\documentstyle[12pt,epsf,cite]{article}

\newcommand{\nc}{\newcommand}
\nc{\beq}{\begin{equation}} \nc{\eeq}{\end{equation}}
\nc{\beqa}{\begin{eqnarray}} \nc{\eeqa}{\end{eqnarray}}
\nc{\lsim}{\begin{array}{c}\,\sim\vspace{-21pt}\\< \end{array}}
\nc{\gsim}{\begin{array}{c}\sim\vspace{-21pt}\\> \end{array}}
\def\Dslash{\not{\hbox{\kern-3pt $D$}}}

\begin{document}

\title{
{\baselineskip -.2in
\vbox{\small\hskip 4in \hbox{hep-th/0511117}}
\vbox{\small\hskip 4in \hbox{LMU-ASC 62/05}}
\vbox{\small\hskip 4in \hbox{TIFR/TH/05-09}}} 
\vskip .4in
\vbox{
{\bf \LARGE Non-Supersymmetric Attractors in String Theory}
}
%\narrowtext
\author{Prasanta K. Tripathy${}^1$\thanks{email: prasanta@theorie.physik.uni-muenchen.de}
~and
Sandip P. Trivedi${}^2$\thanks{email: sandip@theory.tifr.res.in} \\
\normalsize{\it
${}^1$ Arnold-Sommerfeld-Center for Theoretical Physics,} \\
\normalsize{\it Department f\"ur Physik,
Ludwig-Maximilians-Universit\"at M\"unchen,} \\
\normalsize{\it
Theresienstrasse 37, D-80333 M\"unchen, Germany.} \\
\normalsize{\it ${}^2$ Department of Theoretical Physics,}\\
\normalsize{\it Tata Institute of Fundamental Research,} \\
\normalsize{\it Homi Bhabha Road, Mumbai 400 005, India.} 
}}
\maketitle
\begin{abstract}
We find examples of non-supersymmetric attractors in Type II
 string theory compactified on a Calabi Yau three-fold.
For a non-supersymmetric attractor the fixed values  to which  the moduli are drawn at 
the horizon  must minimise an effective potential.
For Type IIA at large volume, we  consider a configuration carrying 
D0, D2, D4 and D6 brane charge. 
When the D6 brane charge is zero, we find for some range of the other charges,
that a non-supersymmetric attractor 
solution exists. When the D6 brane charge is non-zero, we find for some range of charges, 
 a supersymmetry breaking  extremum of the effective potential. 
Closer examination reveals though  that
it  is not a minimum of  the effective potential  and hence the corresponding black hole 
 solution is not an attractor.
Away from large volume, we consider the specific case of the quintic in $CP^4$.
Working in the mirror IIB description we find non-supersymmetric attractors near the Gepner point.

\end{abstract}

\newpage
\section{Introduction and Overview}

Extremal black holes are known to exhibit an interesting phenomenon called the attractor mechanism. 
Moduli fields in these black holes are drawn to fixed values at the horizon.
These fixed values are independent of the asymptotic values for the moduli and are determined entirely by the 
charges of the black hole. So far the 
 attractor mechanism has been mainly discussed  in the context of supersymmetric BPS states.
It was first discovered in \cite{Ferrara:1995ih},  has been studied quite extensively since
then,
\cite{Strominger:1996kf,Ferrara:1996dd,Ferrara:1996um,Ferrara:1997tw,Gibbons:1996af,Denefa,Denef:2001xn},
and has  gained considerable attention lately  due to the conjecture of
\cite{Ooguri:2004zv} and related developments
\cite{LopesCardoso:1998wt,Dabholkar:2004yr,Ooguri:2005vr,Dijkgraaf:2005bp,Sen:2005kj, Kraus:2005vz,
senwa, Kraus:2005zm, Kallosh:2005ax, kalloshb}.
More recently, it was shown that non-supersymmetric extremal  black holes can also exhibit the 
attractor phenomenon \cite{senwa,gijt}.
For some earlier related discussion see also \cite{Gibbons:1996af,Gibbons:1994ff} and especially \cite{Ferrara:1997tw}.
For BPS black holes in ${\cal N}=2$ supersymmetric theories it is known that the
 attractor values minimise the central charge \cite{Ferrara:1996dd}. 
More generally, for supersymmetric and non-supersymmetric extremal black holes,  it was found that 
 the attractor behavior can be understood in terms of an effective potential which depends on the 
charges and the moduli. The fixed values for the moduli are obtained by extremising this potential with 
respect to 
the moduli  and the condition for an attractor is that the resultant extremum is a minimum.

In this paper we  study   examples of non-supersymmetric attractors in string theory.
The setting is  ${\cal N}=2$ supersymmetric compactifications of 
Type II string theory  on a Calabi-Yau manifold \footnote{Our analysis also applies to  
Type II theory on $K3 \times T^2$. In this case the compactification preserves  ${\cal N}=4$ supersymmetry
and in ${\cal N}=2$ language an extra gravitini multiplet is present. As long as fields in this multiplet
 are not excited our results apply.}. 
 We
are  interested in extremal but non-supersymmetric black holes 
in these compactifications. And in this paper we focus on ``big '' black holes with non-zero horizon area classically. 

The discussion  is structured as follows. 
We begin in Section 2,
 by briefly summarising  some of the general  formalism required, with reference  in particular to 
 the effective potential mentioned above. For an ${\cal N}=2$ theory the effective potential only involves 
the vector multiplet moduli, and  can be obtained from the vector multiplet moduli 
space prepotential and the charges carried by the black hole. 
In the cases we encounter in this paper the second derivative matrix 
 of the effective potential has some zero 
eigenvalues. In these situations  higher corrections to the effective potential, beyond quadratic order,
need to be calculated around the extremum. 
We show that the condition for an attractor is that the extremum is a minimum once these 
corrections are included.

Next, in Section 3,  we turn to the specific case of   Type II on Calabi-Yau manifolds. 
In the Type IIA description the vector multiplet 
moduli arise from the (complexified) Kahler moduli of the Calabi-Yau manifold. 
Working self-consistently at large 
volume, we  analyse configurations carrying D6, D4, D2 and D0 brane charge. 
In the analysis we first    consider the case where no D6 branes are present.
In this case we find that for an appropriate ranges of charges a non-supersymmetric attractor exists. 

Next we consider the  case with D6 brane charge. Here we find that for an appropriate range of charges,
an extremum of the effective potential exists and the resulting extremal Reissner Nordstrom black hole, 
obtained by setting the moduli fields at infinity  equal to  their extremum values, breaks supersymmetry. 
However, the effective potential is not a minimum in this case and so the black hole 
is not an attractor. It turns out that the second derivative matrix of the effective potential has 
some zero eigenvalues 
and the leading corrections to the effective potential is cubic along these directions in moduli space. 
This means that generic 
small deviations in the moduli at infinity do not die away near the horizon.  Instead,
they  grow taking the solution further  away from the extremal black hole  as  the horizon 
is approached.  

We have not carried out  an analysis of other possible extrema of the effective potential in the 
case with D6 brane charge. This could reveal the existence of a non-supersymmetric attractor. 
It could also be that the  non-supersymmetric  attractor configuration is   a
 multi-centered black hole \cite{Denef:2001xn}.
We leave  a more complete  investigation along these lines for the future. 

An  important comment about the non-supersymmetric black holes we have analysed  is worth making here. 
The extremum value of the effective potential gives the Beckenstein-Hawking entropy of the black hole. 
One can also compute their entropy from a microscopic point of view.  
For the black holes without any D6 brane charge this can be done using the results of \cite{MSW}.
One finds that the microscopic entropy agrees with the Beckenstein-Hawking entropy.
With D6 brane charge turned on the microscopic counting can be done for the case of $K3 \times T^2$,
as discussed in \cite{ms,hms,kaplan}. 
Once again one finds that the result matches the Beckenstein-Hawking entropy.
This agreement between the microscopic counting and the Beckenstein-Hawking  entropy for non-supersymmetric extremal black holes  is truly striking. 
Note that  with D6 brane charge turned on the black hole is not an attractor, as mentioned above.
Even  so the microscopic and macroscopic answers agree. The agreement of entropy for non-supersymmetric
extremal black holes  has been noticed before, \cite{hms,kaplan,duff1,duff2, dabholkar2,dmr}. 
We hope to develop  a better understanding  for this phenomenon 
 in a forthcoming paper \cite{DSTT}.

In Section 4, we consider an example away from the  large volume limit of  the Calabi-Yau manifold. 
Generally speaking  the analysis is more difficult now, since the effective potential is harder to 
construct explicitly.
One other limit which can sometimes be analysed analytically is that of small complex structure in the mirror IIB 
description. We illustrate this 
 in the case of the mirror quintic. 
The period integrals in this region of moduli space
 can be obtained by a   power series expansion. This allows the effective potential to be constructed 
explicitly. By adjusting  the charges we find examples of non-supersymmetric attractors where the moduli are
fixed self-consistently in the vicinity of Gepner point. It would be interesting to carry out a 
microscopic counting of the entropy in these cases also  to  compare with the gravity answer. 
A similar analysis can be easily repeated for other Calabi-Yau manifolds. Examples  with few moduli,
or where the charges are turned on in a symmetric way so that the minimum lies in a symmetric subspace of the moduli 
space would be most tractable.

\section{Brief Introduction to Non-Supersymmetric Attractors}

\subsection{Brief Introduction}
In this subsection we briefly review the results of \cite{gijt}, see also \cite{senwa}.

We consider a theory whose bosonic terms have the form,
\begin{equation}
  S=\frac{1}{\kappa^{2}}\int d^{4}x\sqrt{-G}(R-2 g_{ij}(\partial\phi^i) (\partial \phi^j)-
  f_{ab}(\phi_i)F^a_{\mu \nu} F^{b \ \mu \nu} -{\textstyle{1 \over 2}} {\tilde f}_{ab}(\phi_i) F^a_{\mu \nu}
  F^b_{\rho \sigma} \epsilon^{\mu \nu \rho \sigma} ).
  \label{actiongen}
\end{equation}
$F^a_{\mu\nu}, a=0, \cdots N$ are gauge fields. 
$\phi^i, i=1, \cdots n$ are scalar fields. The scalars have no potential term but determine the
 gauge coupling constants. 
We note  that $g_{ij}$  refers to the  metric in the moduli space. This is different from the spacetime metric 
\footnote{ 
For ease of discussion,  the moduli space metric was taken to be flat, $g_{ij}=\delta_{ij}$
in eq.(16) of \cite{gijt}, although as discussed after eq.(20) there, the discussion for the attractor goes through 
in the more general case with a non-trivial metric as well.} $g_{\mu\nu}$. 

A spherically symmetric  space-time metric in $3+1$ dimensions takes the form,
\begin{eqnarray}
  ds^{2} & = & -a(r)^{2}dt^{2}+a(r)^{-2}dr^{2}+b(r)^{2}d\Omega^{2}\label{metric2}
\end{eqnarray}

The Bianchi identity and equation of motion for the gauge fields can be solved by a field strength of the form,
\begin{equation}
  \label{fstrenghtgen}
  F^a=f^{ab}(\phi_i)(Q_{eb}-{\tilde f}_{bc}Q^c_m) {1\over b^2} dt\wedge dr + Q_m^a sin \theta  d\theta \wedge d\phi,
\end{equation}
where $Q_m^a, Q_{ea}$ are constants that determine the magnetic and
electric charges carried by the gauge field $F^a$, and $f^{ab}$ is the
inverse of $f_{ab}$. 

The effective potential $V_{eff}$ is then given by,
\begin{equation}
  \label{defpotgen}
  V_{eff}(\phi_i)=f^{ab}(Q_{ea}-{\tilde f}_{ac}Q^c_m)(Q_{eb}- {\tilde f}_{bd}Q^d_m)+f_{ab}Q^a_mQ^b_m.
\end{equation}

It follows  from the Lagrangian, eq.(\ref{actiongen}), and the metric, eq.(\ref{metric2}),
that the scalar fields satisfy the equation,
\begin{equation}
  \label{eomdil}
  \partial_{r}(a^{2}b^{2}g_{ij}\partial_{r}\phi^j)=\frac{\partial_iV_{eff} }{2b^{2}}.
\end{equation}
And the metric components satisfy the relations, 
\begin{eqnarray}
(a^{2}(r)b^{2}(r))^{''} & = & 2, \label{eq1} \\ 
\frac{b^{''}}{b} & = & -g_{ij} \partial_{r}\phi^i \partial_r\phi^j. \label{eq2}
\end{eqnarray}

For the attractor mechanism it is sufficient for two conditions to be met. 
First, for fixed charges, as a function of the moduli, $V_{eff}$ must have a critical point.
Denoting the critical values for the scalars as $\phi^i=\phi^i_0$ we have,
\begin{equation}
  \label{critical}
  \partial_iV_{eff}(\phi_{i0})=0.
\end{equation}
Second, the matrix of second derivatives of the potential at the
critical point,
\begin{equation}
  \label{massmatrix}
  M_{ij}={1\over 2} \partial_i\partial_jV_{eff}(\phi_{i0})
\end{equation}
should have positive eigenvalues. 
  Schematically we can write,
\begin{equation}
  \label{positive}
  M_{ij}>0.
\end{equation}
We will sometimes refer to $M_{ij}$ as the mass matrix and it's eigenvalues as masses
(more correctly $mass^2$ terms) for the fields, $\phi_i$.

Once the two conditions mentioned above are  met  it was argued in \cite{gijt} that the attractor mechanism works. 
There is an extremal Reissner Nordstrom black hole solution in the theory, where the black hole carries the 
charges specified by the parameters, $Q^a_m, Q_{ea},$ and  the moduli take the critical values, $\phi_{i0}$,
 at infinity. 
For small deviations of the moduli from these values at infinity 
 a double-zero horizon extremal black hole solution continues to 
exist. In this extremal black hole the scalars take the same fixed values, $\phi_0$, at the horizon 
independent of their values at infinity.
The resulting horizon radius is given by,
\begin{equation}
  \label{RH}
  b_H^2=V_{eff}(\phi_{i0})
\end{equation}
and the entropy is 
\beq
\label{BH}
S_{BH}={1 \over 4} A = \pi b_H^2. \eeq

In this paper we will encounter  situations where some of the eigenvalues of $M_{ij}$ vanish.
The leading correction to $V_{eff}$ along a  zero-eigenmode direction is then not quadratic but a higher power 
of  the field. Two cases will be encountered, in these the leading power is quartic and cubic 
respectively.  We show below  that as long as the quartic term is 
positive, there is attractor behaviour. In contrast, in the cubic case there is no attractor behaviour.
More generally for attractor behavior the leading correction to the effective potential must be positive.  
We turn to an analysis of these situations  next. 
   
\subsection{Vanishing Mass terms and Attractors}
We begin by  considering   a case  where one  eigenvalue of $M_{ij}$, eq.(\ref{massmatrix}),
 vanishes. We denote the corresponding eigenmode by $\phi$ below.
 The leading
correction to $V(\phi)$ along the $\phi$  direction is then a polynomial,
\beqa
\label{qupot}
V(\phi)& = & V(\phi_0) + \lambda (\delta \phi)^n \\
\delta \phi &  = & \phi -\phi_0.
\eeqa

Our analysis will be along  the lines of \cite{gijt}. We start with an extremal Reissner Nordstrom (eRN)
black hole, with the scalar fields fixed at their extremum values,  and look at small perturbations. The perturbations
satisfy second order equations. The essential point is that if one of the two solutions
of the scalar perturbation equation is well behaved, and vanishes, in the vicinity of the horizon,
then one has attractor behavior. This is because
at infinity the effects of the electromagnetic flux vanish and both solutions to the perturbation equation are
acceptable. Thus, starting with the good solution near the horizon one can extend it to infinity in a smooth fashion.

In the vicinity of the horizon, $r=r_H$,  the metric of the ERN black hole  takes the form,
\beq
\label{metrica}
ds^2=-{(r-r_H)^2\over r_H^2}dt^2 +{r_H^2\over (r-r_H)^2}dr^2 + r^2 d\Omega^2.
\eeq
It is useful to define a coordinate  $t$  given by,
\beq
\label{coord}
e^{-t}\equiv {r-r_H \over r_H}.
\eeq
Note that $t \rightarrow \infty$, as $(r-r_H) \rightarrow 0$.

Let us first consider the case where the eigenvalue of $M_{ij}$ is non vanishing,
and $V(\phi)=V(\phi_0)+{1 \over 2} m^2 (\delta \phi)^2$.
{}From eq.(\ref{eomdil})  the equation for $\delta \phi$ now takes the form \footnote{For simplicity we consider the case where the metric in moduli space is $g_{ij}=\delta_{ij}$. The same conclusions hold more generally.},
\beq
\label{emass}
{\delta {\ddot \phi}} - {\delta {\dot  \phi}} -
{m^2\over r_H^2} {\delta \phi} =0.
\eeq
This corresponds to a particle moving in an upside down harmonic oscillator
 potential with an anti-friction force that aids
in its motion. The attractor solutions correspond to a one parameter family in which $\delta \phi$
reaches the
origin of the potential asymptotically, as $t \rightarrow \infty$.
The approach is exponential in $t$, $\delta \phi=A e^{-\alpha t}$, where, $\alpha$ is determined by the mass,
\beq
\label{valalpha}
\alpha={-1 + \sqrt{1+4 {m^2 \over r_H^2}} \over 2}.
\eeq

Now turn to the case where the mass term vanishes and the leading correction to the potential is quartic,
\beq
\label{quartic}
V(\phi)=V(\phi_0)+\lambda (\delta\phi)^4.
\eeq
For $\lambda >0$ one can see that there is a one parameter family of solutions, which is the
analogue of the slow roll solution in inflation, in which the particle moves due to the anti-friction force
driving it up the hill, with the second derivative term being small. This takes the form,
\beq
\label{ssol}
(\delta \phi)^2={r_H^2 \over 8 \lambda (t +c)},
\eeq
where $c$ is the constant that specifies the one parameter family.
Note that $(\delta \phi) \rightarrow 0$, as $t \rightarrow \infty$, so the attractor value is obtained at the horizon,
 but the approach  is exponentially slower than in the case with non-vanishing mass.

At next order in the perturbation the backreaction on the metric can be calculated. One finds, for the metric,
eq.(\ref{metric2}), using  eq.(\ref{eq1}), eq.(\ref{eq2}),
that,
\begin{eqnarray}
b & = & b_0 +{r_H^3 \over 64 \lambda } {1\over t^2} + \cdots \label{perla} \\
a^2 & = & a_0^2 (1 -  {r_H^2 \over 32 \lambda } {1\over t^2}) +\cdots, \label{perlb}
\end{eqnarray}
where  $a_0^2={(r-r_H)^2\over r_H^2}, b_0=r_H,$ denote the zeroth order near-horizon metric components,
 eq.(\ref{metrica}), and the ellipses indicate terms which are further suppressed in powers of $1/t$.
Since $a_0$ has a double -zero at the horizon, we see from eq.(\ref{perlb}) that after including the backreaction
the metric continues to be a double-zero extremal black hole and from eq.(\ref{perla}) we see that 
the value of the radius $b$ approaches  $r_H$ at the horizon.

So far we have analysed the attractor behavior in the vicinity of the horizon. Extending this analysis
to asymptotic infinity is non-trivial in view of the non-linearity introduced by the quartic term,
eq.(\ref{quartic}).
We have carried out such an analysis numerically and present the results in figure 1.
As one might expect the well behaved attractor solution in the vicinity of the horizon matches smoothly
to an asymptotically flat solution at infinity.

A similar near-horizon  analysis can be repeated in the case where the potential takes the form,
eq.(\ref{qupot}), where $n$ is now a general even power greater than $2$. This leads to the
conclusion mentioned above that the attractor mechanism works as long as
$\lambda>0$ so that the attractor value is a minimum of $V_{eff}$.

In the discussion above we have neglected the mixing between the eigenmode $\phi$ and other massive
 directions in field space.
In general such terms will arise when the potential is expanded about the extremum. However, 
  in the vicinity of the horizon the massive eigenmodes will vanish exponentially more rapidly and such couplings can
 be neglected. 

%\newpage

\begin{figure}
\vskip -.1in
\vglue.1in
\makebox{
\epsfxsize=4.5in
\epsfbox{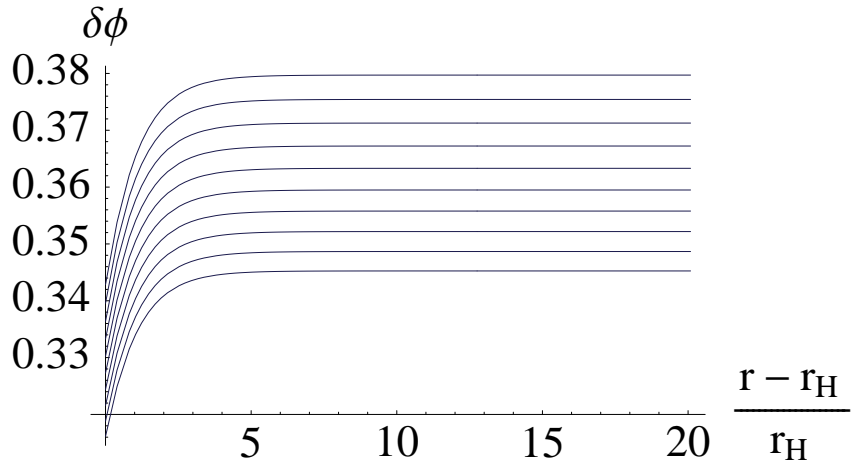}
}
\caption{Plot for the  field $\delta\phi\left({r-r_H\over r_H}\right)$
as a function of $\left({r-r_H\over r_H}\right)$ for different values
of $c$. Here we have chosen $\lambda = 1$ and $r_H = 2$. The integration
constant $c$ takes values $0.1,0.2\cdots 1.0$. The solution is interpolated
from ${r-r_H\over r_H} = e^{-50}$ to ${r-r_H\over r_H} = e^3$. }
\end{figure}

%\newpage

Next, let us consider the case when the power $n$, eq.(\ref{qupot}),  is odd.
For concreteness we take the case when $n=3, \lambda>0$. The equation now takes the form,
\beq
\label{eqcubic}
{\delta {\ddot \phi}} - {\delta {\dot  \phi}} 
-V_{eff}'(\delta \phi) =0.
\eeq
As in earlier cases the second term corresponds to  an anti-friction force while the third term corresponds
to a ''inverted potential'' of the form, $V=-\lambda \phi^3$. 
It is now easy to see that
a small perturbation in the near horizon region,  with $\delta \phi<0$,   does not die away.
Instead, with ${\delta {\dot \phi}}>0$,
both the anti-friction force and the potential drive the perturbation   
 in the same direction towards $\delta \phi \rightarrow 0$. 
As a result the
  perturbation reaches $\delta \phi=0$ at finite $t$ and then continues to grow towards $\delta \phi \rightarrow 
\infty$ as $t \rightarrow \infty$. Once the perturbation becomes large enough the backreaction also becomes significant and our linearised approximation breaks down. 
We have not analysed in full detail the subsequent evolution. It seems reasonable to speculate that
in general  there is no non-singular black hole solution for such a perturbation.
Thus we conclude that for the case where $n=3$ there is no attractor behaviour. 
It is straightforward to see that more generally  a similar conclusion holds for all odd powers of $n$.

A succinct way to summarise our conclusions so far is the following:
there is attractor behaviour   only if 
the effective potential is a positive function of $\delta \phi$. 

So far we have considered the case with  one  zero-eigenvalue of $M_{ij}$. 
In the situations  we will encounter below there are  mutiple zero-eigenvalues. 
One can show that our conclusions above apply in this more general case as well.   Namely, that 
attractor behaviour only holds for a positive 
effective potential. Let us briefly sketch the argument here. 

An equation of the form, eq.(\ref{eqcubic}), now governs each of the zero-eigenmodes. 
If the effective potential is positive, 
a slow-roll solution can be shown to exist in which the second derivative term in eq.(\ref{eqcubic}) is negligible and 
the second term due to anti-friction balances the gradient term from the potential. In this solution 
each zero-eigenmode relaxes to the attractor value as $t \rightarrow \infty$.  
However, if the effective potential is not positive such a solution does not exist in general. 
For appropriately chosen initial values of $\delta \phi_i$ the anti-friction term and the gradient of the potential 
point in the same direction. As a result along some directions  $\delta \phi_i$  vanishes at finite $t$ 
and then continues to increase in magnitude in an unbounded manner till back reaction effects become 
important. 

Two specific situations with zero-eigenvalues   will arise in this paper.
For Type IIA theory at large volume,  without $D6$ brane charge, we will find a positive quartic potential with mutiple zero-eigenvalues in general. 
Once D6-brane charge is included we will find generically a cubic potential again with multiple zero-eigenvalues. 
{}From the discussion above we conclude that  there will be attractor behaviour only in the first case.

\subsection{${\cal N}=2$ Theories}
We conclude this section with a brief discussion of ${\cal N}=2$ supersymmetric theories. 
In these theories, $V_{eff}$ can be expressed, \cite{Ferrara:1997tw}, in terms of a Kahler potential, $K$ and a 
superpotential, $W$ as,
\begin{equation}
  \label{epot}
  V_{eff}=e^K[g^{i \bar j}\nabla_i W (\nabla_j W)^* + |W|^2],
\end{equation}
where $\nabla_iW\equiv \partial_iW+\partial_iK W$. 

The scalars which enter in $V_{eff}$ belong to  vector multiplets and can be described in terms of special geometry. 
Special coordinates, $X^a, a=0, \cdots N$,
 can be chosen  to describe the $N$ dimensional  space.
The Kahler potential and superpotential which appear in eq.(\ref{epot})
  can be expressed in terms of a prepotential $F$ which is
a homogeneous holomorphic function of degree two in these coordinates.  The Kahler potential is given by,
\begin{equation}
\label{kpsg}
K=-\ln Im(\sum_{a=0}^N X^{a*}\partial_aF(X))
\end{equation}
And the superpotential by,
\begin{equation}
\label{superpotsg}
W=\sum_{a=0}^N q_aX^a - p^a \partial_aF.
\end{equation}
Note that in this notation, $q_a$ and $p^a$ are the  parameters, $Q_{ea}$ and
$Q_m^a$ respectively, eq.(\ref{fstrenghtgen}), which determine the electric and magnetic charges of the black
 hole.

For a BPS black hole, the central charge given by,
\begin{equation}
\label{ccharge}
Z=e^{K/2}W,
\end{equation}
is minimised, i.e., 
$\nabla_iZ=\partial_iZ+{1\over 2} \partial_iK Z=0$. This condition is equivalent to,   
\begin{equation}
  \label{attractorsusy}
  \nabla_iW=0. 
\end{equation}
The resulting entropy is given by
\begin{equation}
  \label{susyentropy}
  S_{BH}=\pi e^K |W|^2.
\end{equation}
with the Kahler potential and  superpotential evaluated at the attractor values.

It is worth noting here that in the supersymmetric case, when the central charge is minimised,
\cite{Ferrara:1996dd},  both terms 
on the r.h.s of eq.(\ref{epot}) are  separately minimised.
Since the central charge is minimised, the second term in eq.(\ref{epot}), $|Z|^2$, is at a minimum, and since this 
condition means that $\nabla W$ vanishes the first term in eq.(\ref{epot}) is also at a minimum.
 In contrast, for the non-supersymmetric black hole 
only their sum, $V_{eff}$ is minimised. In particular the central charge is not minimised in the
 non-supersymmetric case.

\section{Type IIA at large Volume}

In this section we analyse  black hole attractors in Type IIA compactifications 
where the  volume of the Calabi Yau manifold is large.
The Calabi-Yau manifold we consider has  $h(1,1)=N$.
The resulting ${\cal N}=2$ low-energy theory has $N$ vector multiplets,
and $N+1$ gauge fields. The one additional gauge field is the graviphoton which lies in the
gravity multiplet.
The leading order prepotential, with no $\alpha^{'}$ corrections, takes the form,
\begin{equation}
\label{prelv}
F=D_{abc}{X^aX^bX^c  \over X^0}
\end{equation}
where $a,b,c=1 , \cdots N$. The intersection numbers $D_{ABC}$ are given by,
\begin{equation}
\label{defdabc}
6D_{abc}=\int_M \alpha_a\wedge \alpha_b \wedge \alpha_c,
\end{equation}
where $M$ denotes the Calabi-Yau manifold and $\alpha_a$ are an integer basis for $H^2(M;{\bf Z})$.

Type IIA theory has D0, D2, D4 and D6 branes. 
D0 branes and D6 branes are electrically and magnetically charged with respect to the graviphoton. 
D2 branes and D4 branes are electrically and magnetically charged with respect to the remaining $N$ gauge fields. 
A general configuration carries charges $(q_0,q_a, p^0, p^a)$, with $a=1, \cdots N$.
To be more specific, an integral basis for $H^2(M;{\bf Z})$, $\alpha_a$, was introduced above.
Let $\Sigma^a$ be a basis of  4-cycles dual to $\alpha_a$.
And let ${\hat \Sigma}_a$ be a basis of 2-cycles Poincare dual to $\Sigma^a$.
Then a D4 brane wrapping the cycle $\Sigma^a$, would    carry  magnetic charge, $p^a$. 
Similarly a D2 brane wrapping the cycle ${\hat \Sigma}_a$  would carry  electric charge $q_a$. 

The special coordinates introduced in subsection 2.3 above correspond to complexified
Kahler moduli of the Calabi-Yau manifold. 
In particular  $J$-  the Kahler two-form  of the Calabi-Yau- satisfies the relation,
\beq
\label{ckm}
\int_{{\hat \Sigma}_a}J=Im({X^a\over X^0}).
\eeq
The  superpotential in supergravity for  a general configuration carrying charges,
 $(q_0,q_a, p^0, p^a)$, 
 is  given by, eq.(\ref{superpotsg}).
 
In the discussion below,   we first consider the case where the D6-brane charge is set to zero.
Subsequently,  we include  D6 branes as well.

\subsection{No D6 branes}
Microscopically, the configuration we consider consists of a D4 brane wrapping the 4-cycle 
$[P]=p_a \Sigma^a$. In addition electric charges, $q_0,q_a$ also arise   due to gauge fields
being turned on,  on  its world volume \footnote{Gravitational Chern-Simons terms will also induce these 
charges.}.

At large volume the prepotential is given  by eq.(\ref{prelv}). 
{}From eq.(\ref{superpotsg}), with $p_0=0$, 
the superpotential takes the form, 
\begin{equation}
\label{IIaspa}
W=q_0X^0 + q_a X^a -3 D_{abc} p^a {X^b X^c \over X^0}.
\end{equation}
And from eq.(\ref{kpsg}) the  Kahler potential is given by 
\begin{equation}
\label{IIAkp}
K=-\ln\left(-iX^0{\bar X}^0 D_{abc}
\left({X^a \over X^0}-{{\bar X}^a \over {\bar X}^0}\right) 
\left({X^b \over X^0} -{ {\bar X}^b\over{\bar X}^0}\right) 
\left({X^c \over X^0}-{{\bar X}^c \over {\bar X}^0}\right) \right)
\end{equation}
The supersymmetric solution for this case are known \cite{lust,shmakova}.

In the subsequent discussion we denote, 
\begin{equation}
\label{defxa}
x^a=X^a/X^0
\end{equation}
and work in the gauge $X^0=1$.

 To begin it is also useful to first set $q_a=0$.
Symmetries suggest the ansatz,  
\begin{equation}
\label{ansatz}
x^a=p^a t.
\end{equation}
Then it is easy to see that the conditions, eq.(\ref{attractorsusy}), are solved by 
\begin{equation}
\label{valt}
t=i\sqrt{q_0/D},
\end{equation}
where 
\begin{equation}
\label{vald}
D=D_{abc}p^ap^bp^c.
\end{equation}

We also introduce for subsequent use the notation,
\begin{equation}
\label{defdadaa}
D_{ab}\equiv D_{abc}p^c, D_a\equiv D_{abc}p^bp^c.
\end{equation}

We are interested in non-supersymmetric solutions to the attractor conditions. As discussed in the previous section
these must minimise the effective potential eq.(\ref{epot}).  It is easy to see that this condition takes the form, 
\begin{equation}
\partial_a V_{eff} =
e^K \left(g^{b\bar c} \nabla_a\nabla_b W \overline{\nabla_cW} + 2 \nabla_a W \overline{W}
+ \partial_a g^{b\bar c} \nabla_b W\overline{\nabla_c W}\right) = 0~.
\label{extpot}
\end{equation}
Symmetries dictate that the ansatz eq.(\ref{ansatz}) must be good in this case as well. 
As discussed in  appendix A.1, substituting for $x^a$ in terms of $t$ from eq.(\ref{ansatz}),
 eq.(\ref{extpot}) takes the
 form, 
\begin{equation}
\label{fform}
{6i \over t} (q_0 -t^2 D) (q_0+t^2 D)=0
\end{equation}

There are two non-singular solutions to eq.(\ref{fform}).  The supersymmetric solution, eq.(\ref{valt})
 is one of them. The second solution is non-supersymmetric and  is given by, 
\begin{equation}
\label{nonsusy}
t=i\sqrt{-q_0/D}
\end{equation}

We note that a non-singular solution requires that the imaginary part of $t$ is non-vanishing 
(otherwise the volume of the Calabi-Yau vanishes) \footnote{There are additional conditions which must be met for the 
supergravity limit to be justified, these are discussed towards the end of this section.}. Thus for a 
given set of charges we can either have a supersymmetric attractor or a non-supersymmetric attractor, depending on whether $q_0/D>0$ or $q_0/D<0$. 

The entropy of the black hole is given by, 
\beq
\label{susyentropytwo}
S=2\pi\sqrt{Dq_0},
\eeq
 in the supersymmetric case, and by
\beq
\label{nonsusyentropy}
S=2\pi\sqrt{-Dq_0},
\eeq
 in the non-supersymmetric case. 
Note that the entropy in the non-supersymmetric case can be  obtained from the supersymmetric case by analytically 
continuing in the charges.   

So far we have set $q_a=0$. This condition is easily relaxed. 
The superpotential, in $X^0=1$ gauge takes the form,
\beq
\label{superpotgen}
W=(q_0 +q_a x^a -3D_{bc}x^bx^c)
\eeq
Since it is quadratic in $x^a$ we can complete the square. 
Let us use the notation $D^{ab}$ for the inverse of the matrix, $D_{ab}$ introduced in eq.(\ref{defdadaa}).
Then, defining the variables, 
\begin{eqnarray}
\label{hatted}
{\hat q}_0 & = & q_0+{1\over 12} D^{ab}q_aq_b \\
{\hat x}^a & =& x^a-{1\over 6} D^{ab}q_b  
\end{eqnarray}
we find the Kahler potential and the superpotential take the same form in terms of the hatted variables as they did 
for the unhatted variables in 
the $q_a=0$ case above. 

The solution in the supersymmetric and non-supersymmetric cases can then be easily written down and  take the form, 
\beq
\label{susygen}
{\hat x}^a=ip^a \sqrt{{\hat q}_0 \over D}
\eeq
and,
\beq
\label{nonsusygen}
{\hat x}^a=ip^a \sqrt{-{\hat q}_0 \over D}
\eeq
respectively. 

And the entropy in the two cases is given by 
\beq
\label{entsusygen}
S=2 \pi \sqrt{D {\hat q}_0},
\eeq
and,
\beq
\label{entnonsusy}
S=2 \pi \sqrt{-D{\hat q}_0}.
\eeq
Note that once again for any set of charges one has either the susy or the non-susy attractor but not both.
One can go from the susy to the non-susy case by reversing  the sign of some charges 
(for example, this can be done by 
taking $p^a \rightarrow  -p^a$ keeping $q_0, q_a$ fixed). 
And analytically continuing in the charges takes the entropy of the susy solution to the non-susy case. 

We have worked in the large volume limit of Type IIA theory above.  The volume  $V$ is determined by the  
 vector multiplet
moduli in IIA theory and is therefore fixed by the attractor mechanism. For the solutions above it is 
given by $V\simeq |{q_0^3 \over D}|^{1/2}$. By taking $q_0$ big enough we can make $V$ big. 
More generally we want the size of all two-cycles and 4-cycles to be big on the string scale.
This leads to the condition,  
$Im(x^a)>>1$. 
We see from eq.(\ref{valt}), eq.(\ref{nonsusy}),
 that this condition can be met by taking  $p^a \sqrt{|q_0/D|} >>1$. 
A similar condition can also be   met by 
by adjusting the charges when $q_a \ne 0$ to ensure that  the Calabi-Yau manifold has large volume. 

It is also worth commenting that the ansatz, eq.(\ref{ansatz}), is singular if the charges $p^a$ are such that $D=D_{abc}p^ap^bp^c=0$. In this case the attractor value for the  volume of the Calabi-Yau  vanishes.
In fact,  this ansatz is inapplicable if $p^a=0$ for any value of the index $a$, since 
 the volume of the   corresponding 2-cycle, ${\hat \Sigma}_a$, vanishes.    

Let us also comment on the physical meaning of the non-supersymmetric solutions we have found. 
We saw in eq.(\ref{susyentropytwo}), eq.(\ref{nonsusyentropy}), 
 that the non-supersymmetric solutions are obtained by say reversing the sign of $q_0$.
Thus starting with a supersymmetric situation containing a D4 brane wrapped on a 4-cycle [P] with induced D0 brane
 charge $q_0$ such that  $q_0/D>0$,  we can get a non-supersymmertric configuration by changing the world volume fluxes
on the brane so that the induced charge $q_0$ reverses sign. The solution we have found above is the supergravity 
description of this microscopic configuration. Similarly with $q_a$ charges 
  also turned on once again changing the sign of the 
D0 brane charge leads to a non-supersymmetric configuration \footnote{One also expects 
 from the spinor conditions that reversing the sign of the D0 brane charge breaks supersymmetry. This is easy to 
verify in a simple case like $K3 \times T^2$.}.   
It is worth commenting that since $|Dq_0|\gg 1$  for a big black hole, reversing the sign of $q_0$ leads to 
$O(1)$ breaking of supersymmetry.
  
So far we have ensured that the attractor values of the moduli extremise the effective potential. 
For an attractor $V_{eff}$ must be minimised. In the supersymmetric case this condition is automatically met,
as was discussed in the previous section. 
In the non-supersymmetric case this needs to be verified by a direct calculation of the second derivatives at the 
extremum. We turn to this next.

\subsection{The Matrix of Second Derivatives}
There are $N$ vector multiplet moduli corresponding to $2N$ real scalars. 
As discussed in appendix B.1 the matrix of second derivatives at the non-susy extremum discussed above has
$N+1$ positive eigen values and $N-1$ zero eigenvalues. 
These zero eigenvalues correspond to the following directions in moduli space. 

Let us write  
\beq
\label{delx}
x^a=p^at + \delta x^a
\eeq
We see from eq.(\ref{nonsusygen}) that at the extremum, where $\delta x^a$ vanishes, $x^a$ is purely imaginary.
The zero mass eigenmodes correspond to $\delta x^a$ being purely real and meeting the condition that 
\beq
\label{condb}
D_{abc}p^bp^c \delta x^a=0
\eeq
 
To analyse the attractor behavior we need to  expand the potential to higher orders in $\delta x^a$. It is enough
for this purpose to only consider the zero eigenmodes, since the other direction have a positive second derivative.
As discussed in appendix B.2 we get keeping terms upto quartic order  that 
\beq
\label{potexp}
V_{eff}=(V_{eff})_0 + e^{K_0}\left[-72{q_0\over D}(D_a\delta x^a)^2 + 36 (D_{ab}\delta x^a\delta x^b)^2 \right]
\eeq
where $(V_{eff})_0,  K_0$ are the values of the effective potential and the Kahler potential at the extremum.  
 
Note that the quartic terms are positive. 
As discussed in section 2.2 this is enough to  ensure that the solution is an attractor.  

This completes our discussion of the Type IIA case with D0, D2 and D4 brane charges turned on. 
We turn to including D6 brane charge next. 

The discussion above goes through essentially unchanged for the case when $q_a\ne 0$, by working in the 
hatted variables, introduced in eq.(\ref{hatted}).

\subsection{Adding D6 Branes}
We now turn to adding $D6$ brane charge.
The configuration we  study has the following  microscopic description. It consists of a single D6 brane 
wrapping the Calabi-Yau $p_0$ times. D4,D2 and D0 brane charges arise due to the world volume gauge field being 
turned on on its world volume (and also due to gravitational Chern-Simons terms).
We will analyse the supergravity description of this configuration below. 
We find  that once again depending on the charges there is either a supersymmetric or non-supersymmetric 
solution which extremises the effective potential. However, somewhat surprisingly, it will turn out that the non-supersymmetric solution is not an attractor. The mass matrix in this case has zero eigenvalues and the leading correction to the potential along these directions of field space is cubic in the perturbation, $
 \delta \phi$, eq.(\ref{cubicattr}). 
For simplicity,  throughout this subsection we restrict ourselves to the  case where $q_a=0$. 

The superpotential in $X^0=1$ gauge takes the form, 
\beq
\label{spd6}
W=(q_0 -3 D_{ab}x^a x^b + p^0 D_{abc}x^ax^bx^c).
\eeq

The susy solution is known \cite{shmakova}, 
\beqa
\label{susyd6}
x^a & = & p^at \\
t & = & {1\over 2D}\left(-p^0q_0 \pm  i \sqrt{q_0(4D -(p^0)^2 q_0)}\right)
\eeqa
where 
\beq
\label{defd}
D=D_{abc}p^ap^bp^c
\eeq

It has entropy, 
\beq
\label{entropysusyd6}
S=\pi \sqrt{4Dq_0 - (p^0)^2 q_0^2}.
\eeq
The supersymmetric solution exists if
\beq
\label{condsusyd6}
{4D\over q_0}>(p^0)^2. 
\eeq

Non-susy extrema of the effective potential $V_{eff}$  are  described in appendix A.2.  
They   exists when ${4D \over q_0}<(p^0)^2$, i.e., when  the inequality, 
eq.(\ref{condsusyd6}), is not met. These  solutions are  somewhat complicated.  
There are two branches depending on whether, ${4D\over q_0}>0$ or ${4D\over q_0}<0$. 
It is useful to define a variable $s >0$ given by, 
\beq
\label{defs}
s=\sqrt{(p^0)^2-{4D\over q_0}}.
\eeq
The   two branches correspond to $|s/p^0|<1$ and $|s/p^0|>1$ respectively. 

The non-susy extrema are obtained by seeking solutions to eq. (\ref{critical})  of the form, 
 eq.(\ref{defxa}), eq.(\ref{ansatz}).
Defining,  
\beq
\label{rit}
t=t_1+it_2, 
\eeq
 one finds that $t_1$ is given by 
\begin{eqnarray}
\label{t1d6}
t_1 = \left\{\matrix{
{2 \over s} {\left(1+{p^0 \over s}\right)^{1/3}-\left(1-{p^0 \over s}\right)^{1/3}\over \left(1+{p^0 \over s}\right)^{4/3}
+\left(1-{p^0 \over s}\right)^{4/3}} &
|{s\over p^0}|>1 \cr
{2\over p^0} {\left(1-{s\over p^0}\right)^{1/3}+\left(1+{s\over p^0}\right)^{1/3}\over \left(1-{s\over p^0}\right)^{4/3}+
\left(1+{s \over p^0}\right)^{4/3}} &
|{s\over p^0}| < 1 \cr
}\right.
\end{eqnarray}
and  $t_2$ by:
\begin{eqnarray}
\label{t2d6}
t_2 = \left\{\matrix{
 {4 s \over (s^2-(p^0)^2)^{1/3} \left((s+p^0)^{4/3}+(s-p^0)^{4/3}\right)} &
|{s \over p^0}|>1 \cr
 {4 s \over ((p^0)^2-s^2)^{1/3} \left((|p^0|+s)^{4/3}+(|p^0|-s)^{4/3}\right)} &
|{s \over p^0}|<1 \cr
}\right.
\end{eqnarray}
In these expressions  the branch cuts are chosen so that all fractional powers are real. 

The entropy of the non-supersymmetric solution is given by 
\beq
\label{entropynonsusyd6}
S=\pi \sqrt{(p^0)^2 q_0^2-4Dq_0}.
\eeq

It is easy to see that the critical  values, eq.(\ref{t1d6}), eq.(\ref{t2d6}), and the entropy go over to eqs.(\ref{entsusygen}) and (\ref{entnonsusy}) of the previous section 
 when $p^0 = 0$. 

For the non-supersymmetric extremum to be an attractor an additional condition must be met.
The extremum  must be a minimum of  the effective potential. 
The matrix of second derivatives in this case is  evaluated in appendix B. 
When $p^0\ne 0$ one finds again that there are $N-1$ zero eigenvalues. 
To decide whether the non-susy solution  is an attractor  one needs to therefore 
expand the potential to higher orders along the zero eigenvalue directions.
This is a rather tedious calculation. Some details are summarised in appendix B.2. 
One finds that the leading corrections to the potential are cubic when $p^0\ne 0$. 
As a result  based on the discussion of section 2.1 we learn,   somewhat surprisingly, that 
the non-supersymmetric  extrema with $p^0\ne 0$ are not attractors.

\section{Mirror Quintic in IIB}
In this section we consider an example away from the large volume limit of IIA theory. 
In general the analysis is more difficult now. One other limit which is sometimes tractable analytically is 
that of small
complex structure in the mirror IIB description. We illustrate this  case 
by studying the example of IIB theory on the mirror
quintic in this section. Our basic strategy will be to consider a black hole with appropriate charges
for which the attractor values of the moduli lie in the vicinity of the Gepner point. Since the period integrals can be obtained in  a power series expansion in this region, an analytic analysis becomes  possible. 
This basic strategy  is analogous to what was done in \cite{GKTT} in the study of 
flux compactifications. 

We begin with some generalities. 
In the IIB theory the vector multiplet moduli correspond to complex structure deformations of the Calabi-Yau manifold.
In general there are $2(h(2,1)+1)$ non-trivial 3-cycles.  A basis of 3-cycles,  $\{A^a,B_a\},$ can be defined 
with,  $A^a \cap  B_b=\delta^a_b, A^a \cap  A^b =0, B_a \cap B_b=0$.
Let $\Omega$ be the holomorphic three-form of the Calabi-Yau manifold. 
Then,
\beqa
\label{periods}
\int_{A^a}\Omega & = & X^a \\
\int_{B_a} \Omega & = & \partial_aF, 
\eeqa
where $X^a$ are the special  coordinates  introduced earlier in our discussion of 
the special geometry of the vector multiplet moduli space and $F$ is the prepotential.

The configuration we consider is obtained by wrapping a D3 brane on a cycle of homology class, 
$[C]=q_a A^a + p^a B_a$. Let $C$ to the cohomology class dual to $[C]$. Now for supersymmetry to be preserved $C$ must 
be a sum of the $(3,0) + (0,3)$ forms 
on the Calabi-Yau manifold.  We will be interested in the supergravity description of the resulting black hole. 
Since the complex structure moduli lie in vector multiplets they will be fixed by the attractor mechanism. 
Depending on the charges, $(q_a,p^a)$,  the resulting complex structure is such that 
sometimes  $C$ will be of   type $(0,3) + (3,0)$ and sometimes it will not. 
In the latter case supersymmetry will be broken. 

The superpotential and Kahler potential can  be expressed as follows:
\beq
\label{spiib}
W  =  \Gamma \cdot \Pi 
\eeq
where $\Gamma$ and $\Pi$ are $2(h(2,1)+1)$ dimensional row and column vectors given by, 
\beqa
\Gamma& = & (-p^a, q_a) \label{defpi}\\
\Pi & = & \left(\matrix{ \partial_aF  \cr  X^a } \right) \label{defpib}. 
\eeqa

The Kahler potential is given by,
\beq
\label{kpiib}
K=-\ln(-i\Pi^\dagger \Sigma \Pi),
\eeq
where the matrix $\Sigma$ is defined in appendix C, eq.(\ref{eqforsig}). 

The mirror quintic is obtained  by starting with the equation
\beq
\label{defmirrorq}
Z_1^5 + Z_2^5 +Z_3^5 +Z_4^5 +Z_5^5 -5 \psi Z_1 Z_2Z_3Z_4Z_5=0
\eeq
in $P^4$, and quotienting by a $(Z_5)^3$ symmetry \cite{candelas}.
It has $h(2,1)=1$,
 so  the vector multiplet moduli space is one-dimensional and the vector $\Pi$ is four-dimensional. 
The complex structure modulus is parametrised by
$\psi$ in eq.(\ref{defmirrorq}).
We will explore solutions to the attractor equations in the vicinity of the Gepner point, $\psi=0$,  below.

To proceed what is needed is  to evaluate the column vector $\Pi$ introduced above in terms of $\psi$. 
As discussed in \cite{candelas} the period integrals of $\Omega$ and thus $\Pi$ can be obtained in a 
power series expansion around $\psi=0$. 
This allows us to write, 
\begin{equation}
\label{msp}
 W = {1\over 25} \left({2\pi i\over 5}\right)^3 \left[ c_0 n\cdot p_0
+ c_1 n\cdot p_1 \psi + c_2 n\cdot p_2 \psi^2 + \cdots \right].
\end{equation}
Here $c_0, c_1, c_2$ are coefficients as defined in appendix C, eq.(\ref{eqforpks}). 
$n=(n_1,n_2,n_3,n_4) $ is a row vector given in terms of the charges by, 
\beq
\label{defn}
(n_1, n_2, n_3, n_4) =5 \Gamma \cdot m, 
\eeq
where $m$ is a matrix defined in appendix C, eq.(\ref{eqform}). 
And $p_0,p_1,p_2$ are column vectors defined in the appendix C, eq.(\ref{eqforpks}). 

The K\"ahler potential and metric are given by, 
\begin{eqnarray}
\label{mkpmet}
&& K = C_0 - \log\left(1 + {c_1^2\over c_0^2} (2-\sqrt{5})|\psi|^2
- {c_2^2\over c_0^2} (2-\sqrt{5})|\psi|^4\right) \cr
&&g^{\psi\bar\psi} = -{c_0^2\over c_1^2 (2-\sqrt{5})}\left(
1 + \left\{ 2(2-\sqrt{5}){c_1^2\over c_0^2} + 4 {c_2^2\over c_1^2}
\right\} |\psi|^2 + \cdots \right).
\end{eqnarray}
The constant $C_0$ is defined in appendix C, eq.(\ref{eqforc0}).

An extremum of the effective potential must satisfy the condition, 
\begin{equation}
\label{miniib}
V'(\psi)=e^K[g^{b\bar c} \nabla_a\nabla_bW \overline{\nabla_cW} + 2 \nabla_aW\overline{W}
+ \partial_ag^{b\bar c} \nabla_bW\overline{\nabla_cW}] = 0.
\end{equation}
We would like to solve this equation self-consistently in the vicinity of $\psi=0$. 

A convenient special case when the algebra simplifies is when $n_1=n_4$ and $n_2=n_3$.  
In this case, as discussed in the appendix C,  we can consistently take 
$\psi$ to also be real. $W$ and $\nabla_\psi W$ are also real then and 
eq.(\ref{miniib})  takes the form,
\beq
\label{simmin}
V'(\psi) = e^K \nabla_\psi W \left(2 W + g^{\psi\bar\psi} \nabla^2_\psi W
+ \partial_\psi g^{\psi\bar\psi} \nabla_\psi W\right) = 0.
\eeq

The susy solution corresponds to setting $\nabla_\psi W=0$. The non-susy solution is obtained from 
\beq
\label{condciib}
2 W + g^{\psi\bar\psi} \nabla^2_\psi W
+ \partial_\psi g^{\psi\bar\psi} \nabla_\psi W=0.
\eeq
From, eq.(\ref{msp}), eq.(\ref{mkpmet}),  for small $\psi$  this takes the form,
\beq
\label{conddiib}
S_1 +S_2 \psi \simeq 0,
\eeq
with, 
\begin{eqnarray}
 S_1 & = & 2 c_0 n\cdot p_0 - {c_0^2\over c_1^2 (2-\sqrt{5})} 2 c_2 n\cdot p_2,
\cr
S_2 & = & 2 c_1 n\cdot p_1 - {6 c_0^2 c_3 \over c_1^2 (2-\sqrt{5})} n\cdot p_3
-  n\cdot p_1 {4 c_0^2 c_2^2\over c_1^3 (2-\sqrt{5})},
\end{eqnarray}
resulting in the solution, 
\beq
\label{mnonsusy}
\psi = - {S_1\over S_2}. 
\eeq

For a solution at small $\psi$ we need to choose charges so that $S_1/S_2 \ll 1$. 
Consistent with our assumption that  $n_1=n_4$ and $n_2=n_3$ it is easy to see that $S_1, S_2$ do not simultaneously vanish. For $S_1$ to vanish we need, 
\beqa
\label{condfiib}
{n_1\over n_2} &  = & {\cos({\pi \over 5})-a \cos({2\pi \over 5}) \over \cos({2\pi \over 5}) 
-a \cos({\pi \over 5})}, \\
a & = & {c_0c_2 \over c_1^2 (\sqrt{5}-2)}.
\eeqa 
This gives, 
\beq
\label{valn}
n_1/n_2\simeq -0.318.
\eeq
Keeping in view the integrality of $n$ this condition is approximately met by taking for example, 
\beq
\label{exn}
n_1=100, n_2=-315.
\eeq
The resulting value of $\psi=1.76\times 10^{-3}$, which is small as expected. 

As discussed in the appendix the matrix of second derivatives has positive eigenvalues in this case. 
Thus the requirements for a non-susy attractor are met. 

Note that the breaking of susy is $O(1)$ in this example  since $|\nabla W|^2/|W|^2 =
(1/(\sqrt{5}-2))(n\cdot p_1/ n\cdot p_0)^2 \sim O(1)$.

The entropy is given by, 
\begin{eqnarray}
S \simeq \pi c_0^2 e^{C_0} \left[|n\cdot p_0|^2 + {1\over \sqrt{5} - 2}
|n\cdot p_1|^2\right] = {2\over \pi^5} \times 10^9 ~. 
\end{eqnarray}

It is worth commenting on   the microscopic configuration  which corresponds to the non-supersymmetric attractor we have found above. The integers eq.(\ref{exn}) with $n_1=n_4, n_3=n_2$ lead to the charges, using eq.(\ref{defn}), 
$(p^a, q_a)=(p^1=400, p^2 = -286, q_1 = 220, q_2 = -143)$ respectively. 
The microscopic configuration is then a D3 branes wrapping the  three-cycle  $[C]=q_a A^a+p^a B_a$. Here we are using 
the basis of 3-cycles introduced at the beginning of this section. For the attractor values of the complex structure
moduli this cycle is non-supersymmetric.

\vspace{.4in}

\noindent
{\Large{\bf{Acknowledgements}}}

\vspace{.2in}
We would like  to thank Gabriel Cardoso,  Kevin Goldstein, Norihiro Iizuka,
Rudra Jena, Dieter L\"ust,  Ashoke Sen, Stephan Stieberger and especially 
Atish Dabholkar for useful discussions. This research 
is supported by the Government of India. S.P.T. acknowledges support from the 
Swarnajayanti Fellowship, DST, Govt. of India. P.K.T. acknowledges support form 
the German Research Foundation (DFG). Most of all we thank the people of India 
for generously supporting research in String Theory. 

\vspace{.4in}
 
\noindent
{\Large{\bf{Appendix}}}

\vspace{.2in}
In appendix  \S A
we present the details on obtaining the nonsusy solutions, first for 
the case without  $D6$ branes and then for the case with  $D6$ branes. 
 In \S B  we compute the matrix of second
derivatives and also expand the potential to higher orders along the zero 
eigen value directions. Finally, in \S C we  discuss the example of 
mirror quintic near the Gepner point. 

\vspace{.3in}
 
\noindent
{\large{\bf A.0~~Nonsusy solutions}}
\vspace{.2in}

We consider type IIA compactification on a Calabi-Yau manifold $M$, with charges
$(q_0,q_a,p^a,p^0)$. 
We denote the vector multiplet moduli 
by $x^a$. Setting the gauge $x^0=1$, we have the superpotential and the 
K\"ahler potential are 
\begin{eqnarray}
W &=& q_0 + q_a x^a - 3 D_{ab} x^a x^b + p^0 D_{abc} x^a x^b x^c \cr
K &=& - \ln\left(-i D_{abc} (x^a-\bar x^a) (x^b-\bar x^b) (x^c-\bar x^c)
\right),
\end{eqnarray}
with $D_{ab}$ given in eq.(\ref{defdadaa}). 
For convenience, we 
introduce the quantities $M_{ab},M_a$ and $M$:
\begin{eqnarray}
&& M_{ab} =  D_{abc} (x^c - \bar x^c) \cr
&& M_a =  D_{abc} (x^b - \bar x^b) (x^c - \bar x^c) \cr
&& M = D_{abc} (x^a - \bar x^a) (x^b - \bar x^b) (x^c - \bar x^c) ~.
\end{eqnarray}
The metric $g_{a\bar b} = \partial_a\partial_{\bar b} K$ can be expressed
in terms of these quantities as 
\begin{eqnarray}
g_{a\bar b} = {3\over M} \left( 2 M_{ab} - {3\over M} M_a M_b\right)~. 
\end{eqnarray}
We also need the inverse of the metric for various computations later.
\begin{eqnarray}
g^{a\bar b} = {M\over 6} \left( M^{ab} - {3\over M} (x^a - \bar x^a)
(x^b - \bar x^b) \right) ~,
\end{eqnarray}
$M^{ab}$ being the inverse of the matrix $M_{ab}$. In what follows, we will 
mainly use the ansatz, 
\begin{equation}
x^a = p^a t = p^a (t_1 + i t_2)~.
\label{ansz}
\end{equation}
 The inverse metric and it's derivative, which
we need in deriving the solutions of the equation of motion, takes the 
following form for the above ansatz,
\begin{eqnarray}
g^{a\bar b} &=& {2 t_2^2\over 3} D \left( {3\over D} p^a p^b - D^{ab}\right) \cr
\partial_a g^{b\bar c} &=& -{i t_2\over 3} D \left( {3\over D} (p^c \delta_a^b
+ p^b \delta^c_a - D^{bc} D_a) + D^{ec} D^{bf} D_{aef}\right) ~.
\label{invmet}
\end{eqnarray}
Here we have used the notation intoduced in eq.(\ref{defdadaa}). 
We now turn to studying non-supersymmetric solutions, first without  D6 branes and then 
with  D6 branes.

\vspace{.3in}
\noindent
{\large{\bf A.1~~No $D6$ branes}}
\vspace{.2in}

In this  case $p^0 = 0$. We will also set $q_a=0$, $a=1, \cdots N$. As explained in \S3, our results 
are also applicable to the case $q_a\ne 0$ 
after  a suitable redefinition for  $q_a, x^a$. 
With $q_a=0$,  the superpotential becomes
\begin{equation}
W = q_0 - 3 D_{ab} x^a x^b ~.
\end{equation}
It is straightforward to find the covariant derivatives of the superpotential.
They have the following form:
\begin{eqnarray}
 \nabla_a W &=&  - 6 D_{ab} x^b - {3 M_a\over M} W \cr % (q_0 - 3 D_{ab} x^a x^b) \cr
 \nabla_a \nabla_b W &=& - 6 D_{ab} + {18 \over M} (M_a D_{bc} + M_b D_{ac}) x^c %\cr &&
- {6\over M} \left( M_{ab}
- {3\over M} M_a M_b\right) W % (q_0 - 3 D_{cd} x^c x^d)
\end{eqnarray}
Since $W$ is a polynomial in even powers of $x^a$, we can set all the 
$x^a$'s to be pure imaginary. The ansatz for $x^a$ then 
becomes $x^a = i p^a t_2 ~.$
The superpotential $W$ and it's covariant derivatives simplifies a lot 
after substituting this ansatz for $x^a$.
\begin{eqnarray}
W &=& (q_0 + 3 t_2^2 D) \cr
\nabla_a W &=& {3 i\over 2 t_2} {D_a \over D} (q_0 - t_2^2 D) \cr
\nabla_a \nabla_b W &=& {3\over 2 t_2^2 D} (q_0 - t_2^2 D) 
\left( D_{ab} - {3\over D} D_a D_a\right)~.
\label{covderv}
\end{eqnarray}
Substituting the expressions for $W$ and it's covariant derivatives from 
eqs.(\ref{covderv}), and using eqs.(\ref{invmet}) in the equations of motion, 
we find 
\begin{equation}
{6 i \over t_2} {D_a \over D} (q_0 - t_2^2 D) (q_0 + t_2^2 D) = 0
\end{equation}
Thus for the nonsusy solution
\begin{equation}
q_0 + D t_2^2 = 0
\end{equation}
and hence 
\begin{equation}
t_2 = \sqrt{- {q_0 \over D}} ~,
\label{nonsusya}
\end{equation}
where as the susy solution corresponds to $t_2 = \sqrt{q_0/D}~.$ {}From this
we see that the nonsusy solution can be obtained from the susy one by setting
$q_0$ to $-q_0$. The susy solution exist for $q_0D>0$ and the nonsusy solution
exists in the region $q_0 D<0$.

\vspace{.3in}
\noindent
{\large{\bf A.2~~Adding $D6$ branes}}
\vspace{.2in}

We will now consider the solutions in presence of $D6$ branes. 
In this case $p^0 \ne 0$ and the superpotential becomes 
\begin{eqnarray}
W = q_0 - 3 D_{ab} x^a x^b + p^0 D_{abc} x^a x^b x^c 
\end{eqnarray}
For later purpose, we summarise the expressions for it's covariant derivatives:
\begin{eqnarray}
\nabla_a W &=& - 3 (2 D_{ab} x^b - p^0 D_{abc} x^b x^c) - {3M_a\over M} W \cr
\nabla_a\nabla_b W &=& - 6 (D_{ab} - p^0 D_{abc} x^c)
 - {6\over M} \left(M_{ab} - 3{M_aM_b\over M}\right) W \cr
&+& {9\over M} \left[M_a (2 D_{bc} x^c - p^0 D_{bcd} x^c x^d)
+ M_b (2 D_{ac} x^c - p^0 D_{acd} x^c x^d)\right]
\end{eqnarray}
For this case, $x^a$ must be a complex quantity with non-vanishing 
real part. We now substitute the ansatz (\ref{ansz}), to obtain the
simplified expressions for $W$ and it's covariant derivatives:
\begin{eqnarray}
W &=& X_1 + i t_2 D Y_1 \cr
\nabla_aW &=& {3\over 2} D_a \left\{ Y_2 + {i\over D t_2} X_2\right\} \cr
\nabla_a\nabla_bW &=& {3\over 2 D t_2^2} \left( D_{ab} - 3 {D_a D_b\over D}\right) X_2
- {9 i\over 2 t_2} \left( D_{ab} -  {D_a D_b\over D}\right) Y_2
\label{covder}
\end{eqnarray}
Here  we have introduced the following 
definitions:
\begin{eqnarray}
\label{defx1toy2}
X_1 &=& q_0 + 3 D t_2^2 (1 - p^0 t_1) - D t_1^2 (3 - p^0 t_1) \cr
X_2 &=& q_0 -  D t_2^2 (1 - p^0 t_1) - D t_1^2 (3 - p^0 t_1) \cr
Y_1 &=& - p^0 t_2^2 - 3 t_1 (2 - p^0 t_1) \cr
Y_2 &=& - p^0 t_2^2 +  t_1 (2 - p^0 t_1). 
\end{eqnarray}
The  susy solution corresponds to 
$X_2 = 0,~Y_2 = 0.$
Solving this for $t_1$ and $t_2$ we obtain
\begin{equation}
t = {1\over 2 D}\left(p^0 q_0 \pm i  \sqrt{q_0 (4 D - (p^0)^2 q_0)}\right)
\end{equation}
This is a valid solution in the range $q_0 (4 D - (p^0)^2 q_0) > 0$, and 
the susy solution ceases to exist beyond this. Thus we expect the nonsusy
solution to occur for  $q_0 (4 D - (p^0)^2 q_0) < 0$. This is indeed the 
case, as we will see below.

Substituting the expressions from eqs.(\ref{covder}) and (\ref{invmet}),
in the equation of motion and equating the real and imaginary parts to zero 
separately, we find
\begin{eqnarray}
\label{eomd6}
X_2 (X_1 + X_2) - Y_2 (D t_2)^2 (Y_1 + Y_2) = 0 \cr
Y_2 (X_1 - X_2) + X_2 (Y_1 - Y_2) = 0
\label{eqmxy}
\end{eqnarray}
We need to solve the above two coupled equations for $t_1$ and $t_2$. To do 
this we will first eliminate $t_1$ from the above two equations to obtain an 
equation for $t_2$ only. We will similarly obtain an equation for $t_1$ by 
eliminating $t_2$ from above. It would then be easier to solve these two 
uncoupled equations.

Eliminating $t_1$ from eqs. (\ref{eqmxy}) we find
\begin{equation}
t_2^2 \left(4 D^2 t_2^2 - q_0 \left(4 D - (p^0)^2 q_0\right) \right) f(t_2) = 0
\end{equation}
with 
\begin{eqnarray}
f(t_2) &=& q_0^3 \left(4 D - (p^0)^2 q_0\right)^3
+ 6 D^2 q_0^2 \left(4 D - (p^0)^2 q_0\right)^2 t_2^2 
+ 9 D^4 q_0 \left(4 D - (p^0)^2 q_0\right) t_2^4
\cr && + D^2 \left(2 D^2 - 4 D (p^0)^2 q_0 + (p^0)^4 q_0^2\right)^2 t_2^6 
\end{eqnarray}
The nonsusy solution for $t_2$ corresponds to $f(t_2) = 0~.$
We can similarly obtain the expression for $t_1$. Eliminating $t_2$ from
eqs.(\ref{eqmxy}), we find 
\begin{equation}
(p^0 q_0 - 2 D t_1) (q_0 - D t_1^2\{3 - p^0 t_1\}) g(t_1) = 0
\end{equation}
where 
\begin{eqnarray}
g(t_1) &=& p^0 q_0^2 - 3 q_0 \left(- 2 D + (p^0)^2 q_0\right) t_1
+ 3 p^0 q_0 \left(- 3 D + (p^0)^2 q_0\right) t_1^2 \cr &&
- \left(2 D^2 - 4 D (p^0)^2 q_0 + (p^0)^4 q_0^2\right) t_1^3 
\end{eqnarray}
We now have a seventh order equation for $t_1$ which has seven solutions.
However it factorizes to two cubic and one linear equation and hence is
exactly solvable.
It is possible to show explicitly that the solution for $t_1$ coming from the 
liner equation or the first of the cubics does not satisfy the equations of 
motion. So we must solve $g(t_1) = 0$ to obtain the nonsusy solution. 

To solve explicitly, let us make the substitution:
\begin{equation}
D = {1\over 4} q_0 \left((p^0)^2 - s^2\right)
\end{equation} 
Here it is worth pointing out that we have taken $D$ to be nonzero throughout.
Thus the solution is meaningful only for $s^2 \neq (p^0)^2$. We will find two 
different solutions in the two regions $s^2 > (p^0)^2 $ and $s^2 < (p^0)^2 $. 

The above substitution in $ g(t_1) = f(t_2) = 0 $ gives 
\begin{eqnarray}
8 p^0 - 12 \left((p^0)^2 + s^2\right) t_1 
+ 6 p^0 \left((p^0)^2 + 3 s^2\right) t_1^2 
- \left((p^0)^4 + 6 (p^0)^2 s^2 + s^4\right) t_1^3 = 0~, \cr
 2^{10} s^6 - 3 ~ 2^7 s^4 \left((p^0)^2 - s^2\right)^2 t_2^2  
+ 36 s^2 \left((p^0)^2 - s^2\right)^4 t_2^4 ~~~~~~~ \cr
- \left((p^0)^2 - s^2\right)^2 \left((p^0)^4 + 6 (p^0)^2 s^2 + s^4\right)^2 t_2^6 = 0  ~.
\end{eqnarray}
The first of the above two equations is a cubic in $t_1$ and hence we can 
solve it exactly. Although the second equation is sixth order in $t_2$, it 
is possible to solve it analytically since it contains only even power in 
$t_2$. Here it is also worth mentioning that not all solutions of the above
two equations actually solve eqs.(\ref{eqmxy}), as is usually the case in 
elimination. Thus we have to carefully and  choose the correct solution. Each
of the cubics allow a pair of complex roots and one real root, and it is 
the real roots which solve eqs.(\ref{eqmxy}). After some simplification, they
take the form eq.(\ref{t1d6}, \ref{t2d6}).  
We can explicitly check that these  expressions for $t_1$ and $t_2$ do
indeed satisfy the equations of motion (\ref{eqmxy}). A simple check shows
that for the above nonsusy solution, the susy breaking scale is $O(1)$. {}From
eqs.(\ref{covder}),(\ref{eqmxy}), we find that 
\begin{eqnarray}
g^{a\bar b} {\nabla_aW \overline{\nabla_bW}\over |W|^2} = 3 ~.
\end{eqnarray}

\vspace{.3in}
\noindent
{\large{\bf B.1~~Diagonalizing Mass Matrix}}
\vspace{.2in}

Here we evaluate the eigenvalues of the mass matrix, eq.(\ref{massmatrix}),
for a configuration consisting of D6, D4, D0 brane charges. We do not include D2 brane charge.
When D6 branes are absent D2 brane charge  can be included in a straightforward manner as was discussed
in section 3.1. 

The matrix elements are given by the double derivative of the effective 
potential. Let  us summarise the required expressions below.  
\begin{eqnarray}
e^{-K_0} \partial_a\partial_d V &=& \left\{ 
g^{b\bar c} \nabla_a\nabla_b\nabla_d W 
+ \partial_a g^{b\bar c} \nabla_b\nabla_d W + \partial_d g^{b\bar c} \nabla_b\nabla_a W
\right\} \overline{\nabla_cW} \cr &+&
 3 \nabla_a\nabla_d W \overline{W}
+ \partial_a\partial_d g^{b\bar c} 
\nabla_b W \overline{\nabla_cW}
- g^{b\bar c}\partial_a g_{d\bar c} \nabla_b W \overline{W} \cr
e^{-K_0} \partial_a\partial_{\bar d} V &=&
g^{b\bar c} \nabla_a\nabla_b W\overline{\nabla_c \nabla_d W}
+ \left\{ 2 |W|^2 + g^{b\bar c}  \nabla_b W\overline{\nabla_c W}\right\} g_{a\bar d}
\cr &+&
 \partial_a g^{b\bar c} \nabla_b W \overline{\nabla_c \nabla_d W}
+ \partial_{\bar d} g^{b\bar c} \nabla_a\nabla_b W \overline{\nabla_c W}
+ 3 \nabla_a W\overline{\nabla_d W}
\cr &+&
  \partial_a\partial_{\bar d} g^{b\bar c} \nabla_b W \overline{\nabla_c W}
\end{eqnarray}
The derivatives are evaluated at the extremum,
\beq
\label{extremumval}
x_0^a=p^a (t_1+it_2),
\eeq
with $t_1,t_2$ given by eq.(\ref{t1d6},\ref{t2d6}). And $K_0$ is the value of the Kahler potential 
at the extremum. 

We now have to use the expression for the superpotential and evaluate 
all the terms. This is a tedious 
but straightforward calculation. We skip some of  the details and give the 
main results below. The terms appearing in the expression for $\partial_a\partial_{\bar d} V$ are given by
\begin{eqnarray}
g^{b\bar c} \nabla_a\nabla_b W \overline{\nabla_c\nabla_dW} &=& - {3 X_2^2\over 2 D t_2^2}
\left( D_{ad} - 9 {D_a D_d\over D}\right) - {27\over 2} D Y_2^2 \left(
D_{ad} - {D_a D_d\over D}\right) \cr
3 \nabla_aW\overline{\nabla_dW} &=& {27\over 4} D_a D_d \left(
Y_2^2 + {X_2^2\over D^2 t_2^2}\right) \cr
2 g_{a\bar d} |W|^2 &=& {3\over D t_2^2} \left({3D_a D_d\over 2 D} - D_{ad}
\right) (X_1^2 + D^2 t_2^2 Y_1^2) \cr
g_{a\bar d} g^{b\bar c} \nabla_bW \overline{\nabla_cW} &=& {9\over 2 D t_2^2}
\left({3D_a D_d\over 2 D} - D_{ad} \right) (X_2^2 + D^2 t_2^2 Y_2^2) \cr
\partial_a g^{b\bar c} \nabla_bW \overline{\nabla_c\nabla_dW} &=& 
{3X_2^2\over D t_2^2} \left(D_{ad} - 3{D_aD_d\over D}\right)
+ 9 D Y_2^2\left(D_{ad} - {D_aD_d\over D}\right) + i {\cal X} \cr
\partial_{\bar d} g^{b\bar c} \overline{\nabla_cW} \nabla_a\nabla_dW &=&
\left(\partial_a g^{b\bar c} \nabla_bW \overline{\nabla_c\nabla_dW} \right)^* \cr
\partial_a\partial_{\bar d} g^{b\bar c} D_bW \overline{D_cW} &=& 
3 D \left(Y_2^2 + {X_2^2\over D^2 t_2^2}\right)
\left({3 D_a D_d \over 2 D} - D_{ad}\right) 
\end{eqnarray}
Here ${\cal X}$ is a real quantity whose explicit expression is not needed
for out purpose.
We can similarly calculate the other matrix elements. Note that 
\begin{eqnarray}
g^{b\bar c} \nabla_a\nabla_b\nabla_d W \overline{\nabla_cW} &=&
{3\over 2} D D_{ad} \left(Y_2 - i {X_2\over D t_2}\right)
\left(6 Y_2 - Y_1 - {i\over D t_2} X_1 + {6 i\over D t_2} X_2\right)
\cr &-&
{9\over 2} D_a D_d \left(Y_2 - i {X_2\over D t_2}\right)
\left(Y_2 + {5i\over D t_2} X_2\right) \cr
\partial_ag^{b\bar c} \nabla_b\nabla_d W \overline{\nabla_cW} &=& 
9 D_aD_d
\left(Y_2 - i {X_2\over D t_2}\right) 
\left(Y_2 + {i\over Dt_2} X_2\right) \cr &-&
 3 D D_{ad} 
\left(Y_2 - i {X_2\over D t_2}\right) 
\left(3 Y_2 + {i\over Dt_2} X_2\right) \cr
3 \nabla_a\nabla_dW\overline{W} &=& - {9\over 2} D D_{ad} 
\left(Y_1 + {i\over Dt_2} X_1\right)
\left( 3 Y_2 + {i\over Dt_2} X_2\right) \cr &+&
 {27\over 2} D_aD_d 
\left(Y_1 + {i\over Dt_2} X_1\right)
\left( Y_2 + {i\over Dt_2} X_2\right) \cr
- g^{b\bar c} \partial_a g_{d\bar c} \nabla_bW\overline{W} &=& 3 D D_{ad}
\left(Y_1 + {i\over Dt_2} X_1\right)
\left(Y_2 + {i\over Dt_2} X_2\right) \cr &-&
 {9\over 2} D_aD_d 
\left(Y_1 + {i\over Dt_2} X_1\right)
\left( Y_2 + {i\over Dt_2} X_2\right) \cr
\partial_a\partial_dg^{b\bar c}\nabla_bW\overline{\nabla_cW} &=&
3 D D_{ad}
\left(Y_2^2 +{X_2^2\over D^2t_2^2}\right)
-{9\over 2} D_a D_d \left(Y_2^2 +{X_2^2\over D^2t_2^2}\right)
\end{eqnarray}
Adding up these we get 
\begin{eqnarray}
e^{-K_0} \partial_a\partial_{\bar d} V
&=& 6 D \left( Y_1^2 + {1\over D^2 t_2^2} X_1^2\right)
\left( 3 {D_a D_d\over D} - D_{ad}\right) \cr
e^{-K_0} \partial_a\partial_d V &=& 6 D D_{ad}
\left({1\over D^2 t_2^2} X_2^2 - Y_2^2 - 2 Y_1 Y_2 - 2 i {(X_1-X_2)\over D t_2}Y_2
\right)
\label{matele}
\end{eqnarray}
Note that, in obtaining the above we have used the equations of motion
(\ref{eqmxy}).

We now set $x^a -x_0^a = y^{1a} + i y^{2a}$ in order to express the mass terms 
in terms of the real fields $y^{1a},y^{2a}~$.
The quadratic terms then take  the form, 
\begin{eqnarray}
S_{mass} & = & 2 \partial_a\partial_{\bar d} V (y^{1a}y^{1d} + y^{2a} y^{2d}) + 2  Re(\partial_a\partial_d V)
(y^{1a}y^{1d} - y^{2a} y^{2d}) \cr && - 2 Im(\partial_a\partial_d V) y^{1a} y^{2d}.
\end{eqnarray}

The mass matrix can then be read off and takes the form,
\beq
\label{massmatrixappb}
M=E \left(3{D_a D_d \over D}-D_{ad}\right) \otimes {\bf I} +  D_{ab} \otimes (A \sigma^3 - B  \sigma^1),
\eeq
where 
\begin{eqnarray}
\label{defeab}
E & = & 12 D e^{K_0} \left(Y_1^2+ {1\over D^2t_2^2} X_1^2\right) \cr
A & =& 12 D e^{K_0} \left({1\over D^2t_2^2}X_2^2-Y_2^2-2Y_1Y_2\right) \cr
B & = & 24  D e^{K_0}  {(X_2-X_1) \over Dt_2}Y_2.
\end{eqnarray}
This is written in  tensor product notation.  Each coordinate $y^{ia}$ has two labels, with $i=1,2$ and $a=1 
\cdots N$. The  ${\bf I}, \sigma_3, \sigma_1$ matrices act in the $2\times 2$ space labelled 
by $i$ and the  $D_{ab}, D_aD_b$ matrices in the $N\times N$ space labelled by $a$. 

To proceed we first diagonalise the $2 \times 2$ matrix
  $A \sigma_3 -B \sigma_1$.
Using the equations of motion, eq.(\ref{eomd6}), we find that the  eigenvalues of this matrix
 are $\pm E$.  
Restricting now to the N dimensional subspace with eigenvalue $+E$ subspace,  $M$ takes the form, 
\beq
\label{restripos}
M_{ab}=3 E{D_a D_d \over D}.
\eeq
It is easy to see that this matrix has $(N-1)$ zero eigenvalues. Any vector $z^a$ with $D_az^a=0$, is a zero 
eigenvector. {}From eq.(\ref{defeab}), we find that 
$E/D= 12 e^{K_0} (Y_1^2+ {1\over D^2t_2^2} X_1^2)>0 $.
 It then follows that the  remaining one eigenvalue is positive.   
Before proceeding let us note  that the zero eigen vectors take the form 
$\left(\matrix{\cos\theta \cr  \sin\theta } \right)$ in the $2\times 2$ subspace, with 
\beq
\label{valtheta}
\tan\theta={B \over A -\sqrt{A^2+B^2}}.
\eeq

Next consider the eigenvector of $A \sigma_3 -B \sigma_1$ with eigenvalue $-E$. Restricting to this 
$N$ dimensional subspace,
$M$ takes the form, 
\beq
\label{restineg}
M_{ab}=2 {E \over D} \left({3 \over 2}  D_a D_d-D_{ad} D\right)
\eeq

Now for a  solution of the form, eq.(\ref{ansz}), the metric, $g_{a {\bar b}}$ becomes, 

\begin{eqnarray}
g_{a\bar b} = {3\over 2 D^2 t_2^2} \left({3\over 2} D_a D_b - D D_{ab} \right)~.
\end{eqnarray}
At a non-singular point in moduli space all $N$ eigenvalues of the metric must be positive. 
Thus we learn that as long as the charges are chosen so that the fixed values for the moduli are at a 
non-singular point in moduli space,  all these $N$ eigenvalues of $M$  are  positive. 

This concludes our discussion of the mass matrix. To summarise, for a general configuration of 
$D6,D4,D0$ brane charges, we find that there are $N-1$ zero eigenvalues and $N+1$ positive eigenvalues of the 
mass matrix. In the case when $D6$ brane charge vanishes, D2 brane charges can also be included in the analysis by 
working in the appropriate ``hatted'' variables, eq.(\ref{hatted}), this means once again the same number of zero and positive eigenvalues. 

\vspace{.3in}
\noindent
{\large{\bf B.2~~Beyond Quadratic Order}}
\vspace{.2in}

Since some eigenvalues   of the mass matrix are zero  we need to calculate terms in the effective potential 
beyond quadratic order  
before deciding whether the non-supersymmetric extremum is an attractor.  These terms need to be calculated 
along the $N-1$ zero eigenvalue directions found above. We turn to this next. 

It is useful to consider the case without any D6 branes first, since the calculations are
 considerably simpler in this case. {}From  eq.(\ref{valt})  we see that $t_1$ vanishes in this case, 
and from eq.(\ref{defx1toy2}) it follows that $Y_2=0$. It then follows  from eq.(\ref{defeab})
that $B=0$ and so we see that the  
zero eigenvectors correspond to $\theta=0$ in eq.(\ref{valtheta}) and are ``purely'' axionic.    
We write, 
\beq
\label{expanda}
x^a = i t_2 p^a + \delta x^a
\eeq
where $\delta x^a$ is real.

We then have 
\begin{eqnarray}
W &=& (q_0 + 3 t_2^2 D - 3 D_{ab}\delta x^a\delta x^b) - 6 i t_2 D_a \delta x^a \cr 
\nabla_b W &=& 4 \delta x^a D t_2^2 g_{b\bar a} + {3 i D_b \over 2 t_2 D}
(q_0 - t_2^2 D - 3 D_{ab} \delta x^a \delta x^b)
\end{eqnarray}
Setting $D t_2^2 = - q_0$ we get 
\begin{eqnarray}
W &=& - ( 2 q_0 + 3 D_{ab} \delta x^a\delta x^b) - 6 i t_2 D_a \delta x^a \cr
\nabla_b W &=& - 4 q_0 \delta x^a g_{b\bar a} + {3 i D_b \over 2 t_2 D}
(2 q_0 - 3 D_{ab} \delta x^a \delta x^b)
\end{eqnarray}

This gives  
\begin{eqnarray}
|W|^2 &=&  ( 2 q_0 + 3 D_{ab} \delta x^a\delta x^b)^2 
+ 36 t_2^2 (D_a \delta x^a)^2 \cr
g^{b\bar c} \nabla_b W \overline{\nabla_cW} 
&=& 16 q_0^2 g_{a\bar b}\delta x^a\delta x^b
+ {9\over 4 D^2 t_2^2} g^{b\bar c} D_b D_c 
(2 q_0 - 3 D_{ab} \delta x^a \delta x^b)^2 \cr
&=& 16 q_0^2 g_{a\bar b}\delta x^a\delta x^b
+ 3 (2 q_0 - 3 D_{ab} \delta x^a \delta x^b)^2  ~,
\end{eqnarray}
from which we obtain
\begin{equation}
V_{eff}(x^a)  
= V_{eff}(it_2p^a) + e^{K_0}\left(
- 72 {q_0\over D}  (D_a \delta x^a)^2 + 36 (D_{ab} \delta x^a \delta x^b)^2 
\right). 
\end{equation}

For the zero eigenvectors, $D_a\delta x^a=0$, so we see as required that the quadratic terms vanish. 
The leading correction to $V_{eff}$ is then quartic. {}From the equation above we see that it's coefficient is 
positive. It then follows from the discussion in Section 2.1   that the  
non-supersymmetric extremum is an attractor. 

Next we turn to the case where the D6 brane charge is non-zero. In this case the calculation of the 
higher order corrections is quite complicated. We omit the tedious details here and simply report the final
 result. Along the zero eigenvector directions we can write, 
$x^a-x^a_0= (\cos\theta + i \sin \theta) \alpha^a$, where $D_a \alpha^a=0$. 
The angle $\theta$ is defined in eq.(\ref{valtheta}).  
One finds that along the zero-eigenvector directions  $V_{eff}$ takes the form, 
\beq
\label{cubicattr}
V_{eff}(x^a)=V_{eff}(x^a_0) + \lambda D_{abc} \alpha^a\alpha^b\alpha^c,
\eeq
where the coefficient $\lambda$ is given by
\begin{eqnarray}
\lambda &=&
{\sin^3\theta\over 2 D^2 t_2^6}\left[
3 D^2 t_2^2|t(p^0t-2)|^2+{5\over 2}|W_0|^2\right] 
- {3\over  t_2^2} |p^0 t - 1|^2 \sin\theta \cr &-&
 {1\over 4 D t_2^3}
 Re\left[
D t (p^0t-2)\left\{
-6 i p^0 t_2 e^{-3i\theta} + 24 p^0 t_2 \sin\theta e^{-2i\theta}
- 12 (p^0\bar t-1) e^{-i\theta} \sin^2\theta\right\} 
\right. \cr &+& \left.
 p^0W_0
\left\{ 6 i \sin\theta e^{-2i\theta} + 4 e^{-3i\theta}\right\}
- 12 p^0 t_2^2 D (p^0 t - 1) e^{-i\theta}\right] ~.
\end{eqnarray}
A straightforward check shows that $\lambda$ does not vanish when $p^0\neq 0$.  
Thus the leading correction to $V_{eff}$ is  cubic.  
This leads to the conclusion that in the case where D6-brane charge is present the non-susy 
extremum we have found is not an attractor.

\vspace{.3in}
\noindent
{\large\bf{C.~~Mirror Quintic}}
\vspace{.3in}

In this appendix, we provide some of the additional formulae used in section 4. 
 We will first calculate the  period vector $\Pi(\psi)$, eq.(\ref{defpib}),
in the vicinity of the Gepner point, $\psi=0$. 
In the Picard-Fuchs basis the period vector ${\bar \omega}(\psi)$  
 can be expressed in terms of a fundamental period: 
\begin{eqnarray}
\omega_0(\psi) = - {1\over 5} \sum_{m=1}^\infty {\alpha^{2m}\Gamma(m/5)
(5\psi)^m\over \Gamma(m)\Gamma^4(1 - m/5)}~,
\end{eqnarray}
as 
\begin{eqnarray}
\bar\omega = - {1\over \psi} \left(2\pi i\over 5\right)^3 \left(
\matrix{\omega_2\cr \omega_1\cr\omega_0\cr\omega_4}\right) ~.
\end{eqnarray}
Here we have done appropriate rescaling of the period vector in order to
keep it nonvanishing at $\psi=0$. The periods $\omega_k(\psi)$ are expressed 
in terms of $\omega_0(\psi)$ as 
\begin{equation}
\omega_k(\psi) = \omega_0(\alpha^k\psi)~,
\end{equation}
with $\alpha = e^{2\pi i/5}$. Now the periods in the integral basis are related
to $\bar\omega$ by
\begin{eqnarray}
\label{valpi}
\Pi(\psi) = m\cdot\bar\omega(\psi)~,
\end{eqnarray}
with 
\begin{eqnarray}
m = \left(\matrix{-{3\over 5} & - {1\over 5} & {21\over 5} & {8\over 5} \cr
0 & 0 & -1 & 0 \cr -1 & 0 & 8 & 3 \cr 0 & 1 & -1 & 0}\right) ~.
\label{eqform}
\end{eqnarray}

For convenience, we introduce the coefficients $c_m$ and vectors $p_m$ as 
follows:
\begin{eqnarray}
&& c_{m-1} = {\Gamma(m/5) 5^m\over \Gamma(m)\Gamma^4(1 - m/5)} \cr
&& p_{m-1} = \left(\matrix{\alpha^{4m}\cr \alpha^{3m}\cr \alpha^{2m}\cr
\alpha^m}\right)
\label{eqforpks}
\end{eqnarray}
{}From the definition of $\alpha$ it follows that, for all values of $m$,
the first and fourth elements in $p_m$ are conjugate to each other and 
so are the second and third elements. It is useful to keep this in mind 
as it will help us later in obtaining the nonsusy solution.

Next,  we  turn to obtaining the superpotential and the K\"ahler potential. 
Let us first express the period vector $\bar\omega(\psi)$ in terms of $c_m$ 
and $p_m$: 
\begin{eqnarray}
\bar\omega = {1\over 5}\left({2\pi i\over 5}\right)^3
(c_0 p_0 + c_1 p_1 \psi + c_2 p_2 \psi^2 + \cdots )
\end{eqnarray}
The periods  in the integral basis may now  be obtained from eq.(\ref{valpi}).
The superpotential  then takes the form eq.(\ref{msp}), 
where  the vector $n=(n_1,n_2,n_3,n_4)$ is defined in eq.(\ref{defn}). 
Similarly we can derive expression for the K\"ahler potential. It is given by eq.(\ref{kpiib}). 
with 
\begin{eqnarray}
\Sigma = \left(\matrix{0&0&1&0\cr 0&0&0&1\cr -1&0&0&0\cr 0&-1&0&0}\right)~.
\label{eqforsig}
\end{eqnarray}
We can substitute the expression for the period vector in the above and  obtain the Kahler potential 
in eq.(\ref{mkpmet}), where
 $C_0$ is an overall additive constant,
\begin{eqnarray}
C_0 = - \log\left[\sqrt{2+2\sqrt{5}}\left({c_0\over 5}\right)^2
\left({2\pi\over 5}\right)^6\right] ~. 
\label{eqforc0}
\end{eqnarray}
It is now straightforward to obtain the metric. We find
 \begin{equation}
g_{\psi\bar\psi} = -(2-\sqrt{5}){c_1^2\over c_0^2}
+\left(2(2-\sqrt{5})^2 {c_1^4\over c_0^4} + 4 (2-\sqrt{5}) {c_2^2\over c_0^2}
\right)|\psi|^2 + \cdots.
\end{equation}
It's inverse is given in  eq.(\ref{mkpmet}).

We now have all the ingredients in hand to find the extrememu of the potential, eq.(\ref{miniib}). 
As discussed in section 4, for simplicity, we restrict to the  
 choice of charges,  $n_1=n_4$ and 
$n_2=n_3$. It is easy to observe that for this choice of charges, the 
product, $n\cdot p_k$, is  always real. Since the coefficients $c_k$ are 
also real, we can consistently choose an ansatz to set $\psi$ to be real.
A quick observation then 
tells that the covariant derivative of $W$ is  also 
real. Thus eq.(\ref{miniib}) can be written in the factorised form, eq.(\ref{simmin}). 

We end by discussing the mass matrix for  the  the non-supersymmetric solution, eq.(\ref{mnonsusy}). 
The potential is given in eq.(\ref{epot}), and we are interested in the second derivatives at the extremum, 
eq.(\ref{mnonsusy}). 
One finds, 
\begin{eqnarray}
\partial_\psi^2 V &=& 6 \left(c_0 c_2 
- {c_0^2 c_3 \over c_1 (2-\sqrt{5})}\right) 
n\cdot p_0 n\cdot p_1 \cr
\partial^2_{\psi\bar\psi} V &=& \left(2 
- {8 c_0^2 c_2^2\over c_1^4 (2-\sqrt{5})}
\right) c_1^2 (n\cdot p_1)^2
- 2 (2-\sqrt{5}) c_1^2 (n\cdot p_0)^2
\end{eqnarray}
For the  choices of $n_1$ and $n_2$, eq.(\ref{exn}), w get, 
$$ \partial_\psi^2 V \sim - 7.8 \times 10^7 ~,~ 
\partial^2_{\psi\bar\psi} V \sim 1.3 \times 10^8 ~, $$
from which it follows that the eigenvalues of the mass matrix 
are positive (with the values $2.08\times 10^8$ and $5.18 \times 10^7$).
%$$\left(\matrix{2.08\times 10^8 & 0 \cr 0 & 5.18 \times 10^7}\right). $$
Thus we see that there are no tachyonic directions and no zero-eigenvalues so that  
 the non-susy extremum is an attractor.

\end{document}